# INVESTIGATION OF THE $K^\pi = 8^-$ ISOMERS IN $N$=74 ISOTONES ON BEAM OF THE WARSAW CYCLOTRON

T. Morek

Nuclear Physics Division, IEP, Warsaw University, Poland

for the Warsaw OSIRIS II Collaboration

The decay of the $K^\pi = 8^-$ isomers in $^{132}$Ce and $^{134}$Nd have been investigated on the beam of the Warsaw Cyclotron using OSIRIS array. Reactions $^{120}$Sn($^{16}$O,4n)$^{132}$Ce and $^{118}$Sn($^{20}$Ne,4n)$^{134}$Nd were used. Two new decay paths have been found in the deexcitation of the $8^-$ isomer in $^{132}$Ce. The hindrance factors for the $E1$, $M2$ and $E3$ transitions deexciting the isomer have been determined. Similar E3 decay of the 410 $\mu$s isomeric state in $^{134m}$Nd has not been observed in our experiment but nevertheless the reduced hindrance factor $f_3 \geq 9$ was determined. The decay properties of the $8^-$ isomers in the $N$=74 isotones are discussed.

## 1. Introduction

The decay modes of $K$-isomers with large changes of the $K$ quantum number are subject to extensive investigations and they are not yet well understood. Isomeric states with $I^\pi$=$8^-$ and $K$=8 are known in all even-even $N$=74 isotones with atomic number $Z$ = 54 - 64 (see refs [4, 5] and references therein). The assignment of a two quasi-neutron ($7/2^+$[404]$\otimes 9/2^-$[514]) configuration is suggested for these isomers. This configuration is supported by the electromagnetic properties of the $8^-$ isomers in $^{128}$Xe and $^{136}$Sm [8, 9]. The respective isomeric half-lives vary from nanoseconds (Xe) to milliseconds (Ce, Ba). It follows from these data that, the eloctromagnetic transitions from $8^-$ isomer to the levels belonging to the ground state and quasi rotational $\gamma$ bands severely violate the $K$ selection rule.The main aim of our work was to extend the experimental information concerning the decay modes of the $8^-$ isomers in $^{132}$Ce and $^{134}$Nd ($N$=74 isotones). The data on the even-even $N$=74 isotones with $Z$=54 - 64 are discused in Sec. 3.

## 2. Experiment

Levels in the $^{132}$Ce nuclei have been populated in the $^{120}$Sn($^{16}$O,4n) reaction at a beam energy of 80 MeV [1]. The $^{134}$Nd nuclei were produced in the $^{118}$Sn($^{20}$Ne,4n) reaction at a beam energy of 100 MeV. The beams were provided by the U200P cyclotron at the Heavy Ion Laboratory of the Warsaw University. The beam had a macro time structure. The macro pulses had a length of 1.0 - 1.5 ms with a time separation adjusted from 3.5 to 28.5 ms. The delayed $\gamma$-radiation was studied with the OSIRIS multidetector array consisted of 6 Compton-suppressed HPGe detectors.

A decay of the 2294 keV, 410$\mu$s isomer in $^{134}$Nd (see Fig. 1) was investigated. This isomer and its decay modes were reported in [2]. The aim of our experiment was to find the $E3$ 596 keV transition expected (in analogy to $^{132}$Ce to deexcite the $K^\pi = 8^-$ isomer to the $5^+$ level from the $\gamma$ band [3]). We are only able to determine the upper limit for intensity of this transition.

The $^{132m}$Ce isomer and its decay mode were observed in [1]. In our investigation (Fig. 1) two new decay paths have been found and the hindrance factors for the $E1$, $M2$ and $E3$ transitions deexciting the isomer have been determined.

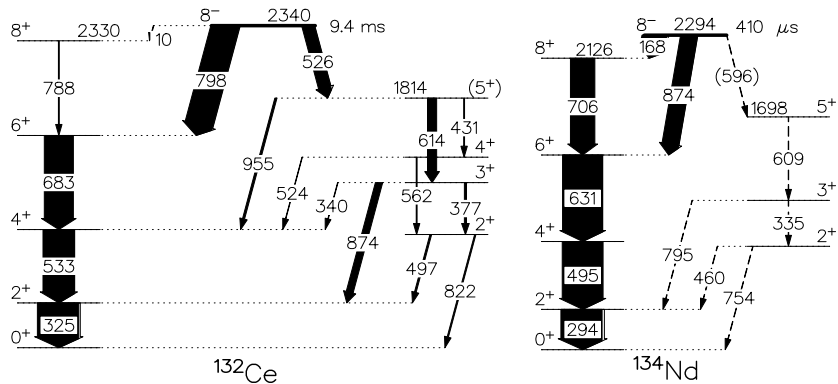

Fig. 1. Decay scheme of the $K^\pi=8^-$ isomers in the $^{132}$Ce and $^{134}$Nd nuclei. Dashed lines denote transitions expected if the $8^-$ isomeric state in $^{134}$Nd decays to $\gamma$ band.

## 3. Discussion

Useful information on the decay properties of the $8^-$ isomer can be obtained from hindrance factors deduced for the deexciting transitions. The

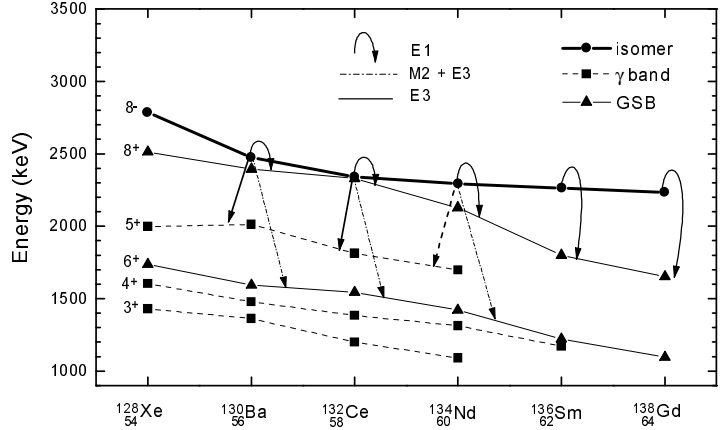

Fig. 2. Systematics of the energy levels and isomeric transitions in $N$=74 nuclei. Decay modes of the $8^-$ isomers are also shown. Dashed line denotes expected E3 transition in $^{134}$Nd

reduced hindrance factor $f_\nu$ of a $\gamma$-transition is defined as

$$f_\nu = (T^p_{1/2}/T^W_{1/2})^{1/\nu},$$

where $T^p_{1/2}$ is the partial half-life of the $\gamma$-transition, $T^W_{1/2}$ is the corresponding Weisskopf single particle estimate, $\nu$ is the degree of $K$-forbiddenness defined as $\nu = \Delta K - \lambda$, where $\lambda$ is the multipolarity of the radiation. In the case of $E1$ transitions, $T^W_{1/2}$ was multiplied by a factor of $10^4$ to take into account systematics of the $E1$ hindrance factors. The reduced hindrance factors for the $\gamma$-ray transitions deexciting the $K^\pi$=$8^-$ isomeric states in the even-even $N$=74 isotones are presented in Fig. 3. The dependence of the $f_\nu$ values on the atomic number $Z$ can be used as a source of information about the mechanism of weakening of $K$-forbiddeness.

Three types of the $K$-forbidden transitions are observed in discussed nuclei:

- $E1$ transitions between $(I^\pi,K$=$8^-,8) \to (I^\pi,K$=$8^+,0)$ states; $\nu = 7$.
- $M2$ transitions between $(I^\pi,K$=$8^-,8) \to (I^\pi,K$=$6^+,0)$ states; $\nu = 6$.

For $^{132}$Ce and $^{134}$Nd the $E3/M2$ mixing ratio is not known, therefore the experimental points indicate only lower limits of $f_6$. For $^{130}$Ba the $E3$ admixture to the $M2$ multipolarity was deduced from experimental value of electron conversion coefficient [4, 6].

- $E3$ transitions between $(I^\pi, K=8^-, 8) \to (I^\pi, K=5_\gamma^+, 2)$ states; $\nu = 3$. In this case we assumed, that the observed transitions have pure E3 character, since a significant admixture of $M4$ multipolarity is unlikely.

The $K^\pi = 8^-$ isomers in $^{130}$Ba, $^{132}$Ce, $^{134}$Nd, $^{13}$Sm and $^{138}$Gd isotone decay via forbidden $E1$ transitions with a degree of $K$ forbiddeness $\nu = 7$ to $8^+$ members of the ground state band. The $E1$ transition rates and respective reduced hindrance factors $f_7$ (circles in fig. 3) vary significantly from isotone to isotone Recently , in Ref. [5] deexcitation mechanism through $8^- \to 8^+$, $E1$ transitions has been suggested for the $N=74$ isotones. The proposed mechanism involves the interaction between the gsb and the s-band. The admixture of the s-band wave function with high value of $K$ to the wave functions of the gsb members depends on the interaction strength between these two bands. The interaction strenght can be evaluated from the experimental alignment plot. It allows to calculate relative values of reduced hindrance factors ($f_7$) $^{130}$Ba, $^{132}$Ce, $^{134}$Nd, $^{136}$Sm and $^{138}$Gd [5]. These values normalized to the experimental $f_7$ value for $^{130}$Ba are presented in Fig. 3 as a solid line. The experimental $f_7$ values (among them our point for $^{132m}$Ce) agrees very well with the calculated ones.

The isomer decay branch which leads via $M2(+E3)$ transition with $\nu = 6$ to the $6^+$ members of the ground state band is only known in $^{130}$Ba, $^{132}$Ce and $^{134}$Nd. The corresponding reduced hindrance factors $f_6$ are marked by squares in fig. 3.

One can try to use the same mechanism as proposed above to explain the values of the reduced hindrance factors $f_6$ for $M2$ transition from the $8^-$ isomers to the $6^+$ members of the yrast band. Admixtures of the s-band into the $6^+$ yrast state wave functions have been calculated [5, 12]. The resulting $f_6$ values normalized to the experimental $f_6$ value for $^{130}$Ba, are shown in Fig. 3 (dashed curve). The experimental value for $^{134}$Nd differs significantly from the calculated one. This may indicate that band-mixing model is not sufficient to explain the data for the $M2$ $(8^- \to 6^+)$ transitions.

The $f_3$ data for the $E3$ transitions from the isomeric $8^-$ state to the $5_\gamma^+$ ($K=2$) state are very scarce (see Fig. 3). However, one may argue that a non-axial deformation may be responsible for the weakening of the $K$-forbiddeness in the case of the isomeric decay to the quasi-rotational $\gamma$-band members. For the deduced value of the deformation parameter $\gamma \approx 24°$ the wave function of the $5_\gamma^+$ state calculated in the framework of the Davydov-Filippov model [13] contains about 4% of $K=4$ admixture to the $K=2$ wave function in the nuclei of interest. This may facilitate the $8^- \to 5^+$, $E3$ transition observed in $^{130}$Ba and $^{132}$Ce through $K=7$ and $K=4$ admixture to the wave function of the initial and final states, respectively. Such $K=7$ admixture coming from the $7/2^+[404] \otimes 7/2^-[523]$ two-neutron configuration in the $8^-$ isomer was found in $^{134}$Nd [10]. Similar values for the amplitude of $K=4$ admixture have been deduced for both nuclei (since the $\gamma$-deformations are similar) explaining the nearly equal $f_3$ values for these transitions. In $^{134}$Nd similar E3 decay of the 410 $\mu$s isomeric state has not been observed. However, we were able to determine that in this case the reduced hindrance factor $f_3 \geq 9.5$. Both hindrance factors: $f_3$ and $f_6$ suggest sharp difference in the $^{134}$Nd isomer structure in comparison with $^{130}$Ba and $^{132}$Ce.

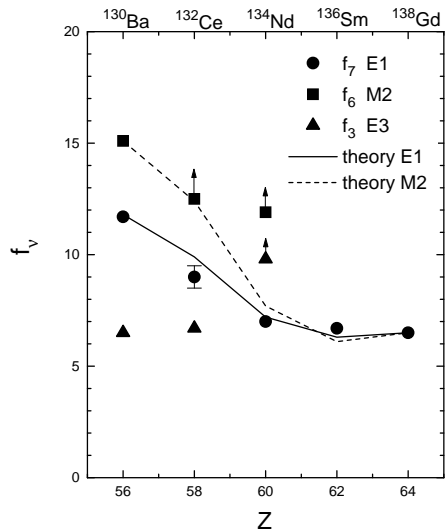

Fig. 3. Systematics of the reduced hindrance factors for the even N=74 isotones. Calculated values (solid and dashed lines) for $f_6$ anf $f_7$ are also shown. In the case of $E1$ transitions the Weisskopf estimates used to calculations of $f_7$ are multipied by factor of $10^4$ to take into account their generally higher hindrance.

## 4. Conclusions

The decay properties of the isomeric $K^\pi=8^-$ states in the $^{132}$Ce and $^{134}$Nd nuclei have been studied in the experiment. The isomers decay via highly $K$-forbidden $\gamma$-transitions to the members of the ground state band and quasi rotational $\gamma$-band. In $^{132}$Ce the reduced hindrance factors $f_7 = 9.0(0.5)$, $f_3= 6.7(0.1)$ and the lower limit $f_6 \geq 12.5$ fit nicely into the systematics of the hindrance factors for the even-even $N=74$ isotones. A simple two-band mixing model involving an interaction of the gsb with s-band, as suggested in ref. [5], allows for an explanation of the observed Z dependence of the reduced hindrance factors $f_7$ for E1 transitions. However this model fails to reproduce the $f_6$ values. In the case of the $E3$ transitions (reduced hindrance factors $f_3$) it is shown that the nonaxial deformation should be taken into account. The similar values of $f_3$ for $^{130}$Ba and $^{132}$Ce may be related to a nearly constant $\gamma$-deformation deduced for these nuclei. The E3 decay of the isomeric state in $^{134m}$Nd has not been observed. However, we were able to determine that in this case the reduced hindrance factor $f_3 \geq 9.5$. A more detailed study of the $K^\pi=8^-$ isomeric decay in the heavier $^{136}$Sm and $^{138}$Gd nuclei would be very helpful for a better understanding of the mechanism of $K^-$ forbidden $\gamma$-transitions.

Warsaw OSIRIS II Collaboration : T. Morek, J. Srebrny, Ch.Droste, M. Kowalczyk, T. Rząca-Urban, K. Starosta - *Nuclear Physics Division, IEP, Warsaw University, Poland*; W. Urban - *Nuclear Spectroscopy Division, IEP, Warsaw University, Poland*; R. Kaczarowski, E. Ruchowska - *The Andrzej Sołtan Institute for Nuclear Studies, Świerk, Poland*; M. Kisieliński, A. Kordyasz, J. Kownacki, M. Palacz, E. Wesołowski, M. Wolińska - *Heavy Ion Laboratory, Warsaw University, Poland*; W. Gast, R.M. Lieder - *Institut für Kernphysik, Forschungszentrum Jülich, Germany*; P. Bednarczyk, W. Męczyński, J. Styczeń - *The Niewodniczański Institute of Nuclear Physics, Kraków,Poland*

## REFERENCES

[1] T. Morek et. al, Phys.Rev **C63** 034302 (2001)
[2] D.G. Parkinson et al. Nucl.Phys. **A194** (1972) 443
[3] K.S. Vierinen et al. , Nucl.Phys. **A499** (1989) 1
[4] R.B. Firestone, Table of Isotopes, $8^{th}$ Edition, 1996
[5] A.M. Bruce,A.P. Byrne, G.D. Dracoulis, W. Geletly, T. Kibedi, F.G.Kondev, C.S. Purry, P.H. Regan, C. Thwaites and P.M. Walker, Phys.Rev. **C55**, 620 (1997)
[6] H. Rotter, K.F. Alexander, Ch. Droste, T. Morek, W. Neubert and S. Chojnacki, Nucl.Phys. **A133**, 648 (1969)
[7] F.R. Xu, P.M. Walker and R. Wyss, Phys.Rev. **C59**, 731 (1999)
[8] T. Lönnroth, S. Vajda, O.C. Kistner and M.H. Rafailovich, Z.Phys. **A317**, 215 (1984)
[9] P.H. Regan, G.D. Dracoulis, A.P. Byrne, G.L. Lane, T.Kibedi, P.M. Walker and A.M. Bruce, Phys.Rev. **C51**, 1745 (1995)
[10] C.M. Petrache et al., Nucl.Phys. **A617**, 249 (1997)
[11] J. Meyer-ter-Vehn, Nucl.Phys. **A249**, 141 (1975)
[12] A.M. Bruce, private communication
[13] A.S. Davydov and G.F. Filippov Nucl.Phys. **8**, 237 (1958)